\documentclass[showpacs,aps,preprint,amsmath,amssymb,feynmp]{revtex4}

\usepackage{graphicx}
\usepackage{dcolumn}
\usepackage{bm}
\usepackage{feynmp}

\def\nablab{{\mbox{\boldmath $\nabla$}}}
\def\thetab{{\mbox{\boldmath $\theta$}}}
\def\fsz{\footnotesize}
\def\comment#1{}

\def\gtsim{{\,\,\raisebox{-1mm}{${\displaystyle{\mathop{\sim}^{>}}}$}\,}}
\newcommand{\nc}{\newcommand}
\nc{\beq}{\begin{eqnarray}}
\nc{\eeq}{\end{eqnarray}}
\nc{\scs}{\scriptstyle}
\nc{\setval}{\fmfset{wiggly_len}{3mm} \fmfset{arrow_len}{1.5mm}
    \fmfset{arrow_ang}{13} \fmfset{dash_len}{1.5mm}\fmfpen{0.125mm}
    \fmfset{dot_size}{2thick}}
\begin{document}
\title{Charged fixed point in the Ginzburg-Landau 
superconductor and the role of the 
Ginzburg parameter $\kappa$}
\author{Hagen Kleinert and  Flavio S. Nogueira}
\affiliation{Institut f\"ur Theoretische Physik,
Freie Universit\"at Berlin, Arnimallee 14, D-14195 Berlin, Germany}

\date{Received \today}

\begin{abstract}
We present a
semi-perturbative approach which
yields an
infrared-stable fixed point
in the Ginzburg-Landau for $N=2$, where $N/2$ is the 
number of complex components.
The calculations are done
in $d=3$ dimensions
and below $T_c$,
where
the renormalization group functions can be expressed
directly as
functions
of the Ginzburg parameter $\kappa$ which is
the ratio between
the two fundamental scales of the problem, the penetration depth
$\lambda$ and the correlation length $\xi$.
We find a charged fixed
point for $\kappa>1/\sqrt{2}$, that is, in the type
II regime, where $\Delta\kappa\equiv\kappa-1/\sqrt{2}$ is
shown to be
a natural expansion parameter.
 This parameter
controls a
 momentum space instability in the
two-point
correlation function of the
order field.
This instability appears
at a nonzero  wave-vector ${\bf p}_0$ whose magnitude
scales like $\sim\Delta\kappa^{\bar{\beta}}$, with a critical
exponent $\bar{\beta}=1/2$
in the one-loop approximation, a behavior
known from
magnetic systems with a Lifshitz point in the phase diagram.
This momentum space instability
is argued to be the origin
of the
 negative $\eta$-exponent of the order field.  \\
\end{abstract}

\pacs{74.20.-z, 05.10Cc, 11.25.Hf}
\maketitle
\section{Introduction}
Whereas the field-theoretic
understanding of the universal critical
properties
of the superfluid phase transition
has become
quite satisfactory \cite{ZJ,HKSF},
the
renormalization group (RG)
have so far not been completely successful 
for the
superconducting phase transition.
The standard field theory
investigated
in this context
is the
 Ginzburg-Landau (GL) model, initially
in $4-\epsilon$ dimensions by
 Halperin, Lubensky, and Ma (HLM)\cite{HLM1},
who
generalized a
calculation of Coleman and Weinberg \cite{CW} in
four dimensions, and found no infrared stable fixed point
at
the one-loop level.
The renormalization group (RG) flow
was shown
to run off to infinity in the $g-f$-plane if 
$N<365.9$, with $N/2$ being the number of complex 
components of the scalar order parameter field. 
Here $g$ and $f$ are
the dimensionless versions of the quartic self-coupling
$u$ of the order parameter and the square charge $e^2$,
respectively.
HLM
interpreted this result as indicating
that the phase transition in a superconductor - 
which corresponds to a order parameter with $N=2$ -  
is of first-order.
They supported this scenario by a mean-field
estimate
of the
effective potential obtained by
integrating out
 the vector potential, which led
to a
term \cite{Schakel}
$\sim|\psi|^{4-\epsilon}$,
which leads to first-order phase transition.
This was in contrast to Monte Carlo
simulations of lattice superconductors  in 1981
by
Dasgupta and Halperin
\cite{Dasgupta} where
a clear second-order phase transition was found.
It also contradicted work by Lawrie and Athorne \cite{LawAth}
who performed a RG study in a non-linear realization of the GL model
and found no indication
 of a first-order phase transition.
Moreover, the first-order scenario was also contradicted by
experiments on the
 smectic-A to nematic  phase transition in liquid crystals \cite{AN}
which according to de Gennes \cite{deGennes} can also
 be described by
a GL model.

For small $e^2$
 or large Ginzburg parameter $ \kappa $,
the GL-model
lies in the deep type II regime
where it
goes over into
a pure complex $|\psi|^4$-model which describes
the  superfluid phase transition,
which is the best-understood
 second-order transition experimentally and theoretically.
We may therefore expect a large-$ \kappa $
portion of the type II regime
to undergo a second-order transition.
Indeed,
the Monte
Carlo simulations of Ref. \onlinecite{Dasgupta} and the
non-linear realization of Ref. \onlinecite{LawAth}
were both
in this regime.

The arguments leading to a first-order transition
could therefore at best be reliable in the large-$e^2$
regime
where the superconductor is of type I and the Ginzburg parameter
$ \kappa $ is small. Only in this regime
can the $\epsilon$-expansion
 be trusted
and the important problem arose
how to
find a theory explaining both type I and type II
regimes.

One solution to this problem
was proposed in 1982 by
one of us\cite{Kleinert,KleinertBook} 
by deriving
a disorder field theory dual to the original GL
model.
The RG analysis of the disorder
field theory shows very clearly that a charged fixed point
exist in the original GL model \cite{Kiometzis2}.
The
disorder field theory exhibited
a first- and a second-order regime
with a tricritical point
at\cite{Kleinert} $ \kappa _t=(3/\pi) \sqrt{3/2}\left[1-(4/9)^4 \right]\equiv
0.79/ \sqrt{2}
  $, a result
confirmed only recently with excellent agreement
by elaborate Monte Carlo simulations \cite{SUD} which gave
$ \kappa _t=(0.76\pm0.04)/ \sqrt{2}$.
Since the disorder theory
is in  pure $|\phi|^4$ universality class,
the
 critical exponent $\nu$ of the disorder field
in the second-order regime
has  the $XY$-value $ \nu \approx0.67$ \cite{Kiometzis2},
a result confirmed  by Monte Carlo simulations
\cite{Olsson}.

The RG situation of
the original GL
model, however, remained completely unclear.
No fixed point at $N=2$ appeared in an extension to two-loops
by Tessmann \cite{Tess},
a result confirmed later by Kolnberger and Folk using a more 
thorough analysis 
\cite{Kolnberger,NoteTess}.
Folk and Holovatch
\cite{Folk} managed to find
a charged fixed point
only by setting up a suitable
Pad\'e-Borel resummation of the $\epsilon$-expansion.
A three-loop calculation using the $\epsilon$-expansion is
presently under progress. As a first step,  the ground state energy
of the GL model was computed recently
up to three-loop by
 Kastening, Kleinert, and Van den Bossche
  \cite{Kast}.

Another method that leads to a charged fixed point at $N=2$ 
was used by Bergerhof {\it et al.} \cite{Berg}. Their
method is based in the so called {\it exact renormalization group} 
\cite{ERG}. In this approach, exact renormalization group equations 
are solved approximately 
using a truncated form for a scale-dependent effective 
action. The problem with this type of approach is that it 
seems to be uncontrolled, 
that is, there is no obvious expansion parameter. But this seems to 
be a common feature of most fixed dimension approaches. 
However, as we shall see 
later on in this paper, fixed dimension approaches are the best 
candidates to solve this old problem.  

At fixed dimension $d=3$ there is the approach by 
Herbut and Tesanovic\cite{Herbut} who  
performed a
 perturbative calculation directly
at the
critical point. These authors
 obtained a charged fixed point at one loop
and $N=2$ by using
 two different momentum scales and
different renormalization
points for the two dimensionless couplings,
$f\equiv e^2/\mu'{}^{4-d}$ and
$g\equiv u/\mu^{4-d} $.
This introduces
 a new  parameter $c\equiv\mu/\mu'$.
They find charged fixed points
if $c>5.16$. Their
 approach has the problem
that the parameter $c$ must
be fixed from the outside. They
do this by
demanding that
the value of the ratio
$\kappa=\sqrt{g/2f}$ matches
 at the  tricritical point
with Kleinert's
value\cite{Kleinert}
$\kappa_t\simeq 0.79/\sqrt{2}$,
and found
a critical exponent
 $\nu\approx 0.53$.
The authors used also
an older
Monte Carlo estimate\cite{Barth} $\kappa_t\simeq 0.42/\sqrt{2}$
to fix $c$ which led
to a $ \nu $ closer to the the $XY$ value.
In the light of the recent simulation work \cite{SUD}
obtaining the value $\kappa_t\simeq 0.77/\sqrt{2}$,
the result $\kappa_t\simeq 0.42/\sqrt{2}$ becomes obsolete.

There are two important  universal  observations made in
Ref. \onlinecite{Herbut}:
 first, if a charged fixed point exists, then the
 anomalous dimension
of the vector potential is {\it exactly}
given by
 $\eta_A=4-d$ for $d\in(2,4]$. Second, this anomalous dimension
implies the existence of \cite{Herbut,Olsson}
a {\it dimension-independent} scaling relation
$\lambda\propto\xi$
between
the magnetic penetration depth $ \lambda $ and
the coherence length
$\xi$. Subsequently, the AC conductivity
$\sigma(\omega)$  was shown\cite{Nogueira2,Mou}
to have the scaling behavior
$\sigma(\omega)\sim\xi^{z-2}$, where $z$ is the dynamic exponent.
These results
are in contrast to the {\it dimension-dependent}
scaling relations
$\lambda\sim\xi^{(d-2)/2}$ and
$\sigma(\omega)\sim\xi^{2-d+z}$ derived by
D.S. Fisher {\em et al.\/}\cite{Fisher}
by {\em neglecting\/}
 thermal  fluctuations of the magnetic field, leading to
$\eta_A=0$.

It is important to emphasize
  that
in Ref.~\onlinecite{Herbut} the
two renormalization scales $\mu$ and $\mu'$
play a decisive role for the emergence
of the charged fixed point.
However, since
the correlation functions
were calculated
{\em at\/} the critical point,
there was no direct relation between
     $\mu$ and $\mu'$
 and the two physical length scales
of the GL model. One of this lengths, the penetration depth, 
is observable only in the
 {\it ordered}, since the vector potential becomes massive 
only below $T_c$ through the Anderson-Higgs mechanism. 
It is therefore desirable to study the
RG flow in such a way that $T_c$ is approached from 
below, and this is what we shall do in this paper by employing a 
semi-perturbative approach.  
 
Since below $T_c$ 
 the vector potential is a massive field, it 
provides us with
the second physical mass scale of the system, and thus with the Ginzburg
parameter $ \kappa $
on which the RG functions depend explicitly. 
We shall exhibit a charged fixed point for $N=2$ and 
$\kappa>1/\sqrt{2}$, that is, inside the type II regime.
In our analysis, $\Delta\kappa\equiv\kappa-1/\sqrt{2}$ appears as
a natural expansion parameter when $T_c$ is approached from 
below. This is related to the existence of a singular behavior at 
$\kappa=1/\sqrt{2}$. This singular behavior is already apparent 
in the mean field solution in a uniform external magnetic field 
by Abrikosov \cite{Abrikosov}. For example, the magnetization 
diverges at $\kappa=1/\sqrt{2}$. 
To our knowledge, in the RG context this singular 
behavior in $\kappa$ has never been explored before. We shall show 
that the existence of the charged fixed point is related to 
a singular behavior at $\kappa=1/\sqrt{2}$ occuring in the 
vector potential correlation function. This singularity has 
the physical meaning that it represents the point of separation 
between type I and type II superconductivity. As discussed in 
Ref. \onlinecite{SUD}, the value of $\kappa$ separating these two 
regimes should coincide with $\kappa_t$, being therefore lower 
than $1/\sqrt{2}$. The approximation we consider here is not able 
to correct upon the value $\kappa=1/\sqrt{2}$, but higher order 
corrections will do. At this point it is worth to explain what we 
mean by higher order. Our method is semi-perturbative. By this 
we mean that we start by assuming that a perturbative expansion in $f$ 
and $g$ is 
possible. However, we parametrize our theory 
by $f$ and $\kappa$ such that all RG 
functions depend on these parameters. This is possible because 
$\kappa^2=m^2/m_A^2=g/2f$, where $m=\xi^{-1}$ is the Higgs mass 
and $m_A=\lambda^{-1}$ is the vector potential mass. 
We use $m$ as the running scale for the RG.
Thus, 
$\kappa$ arises in the RG functions from two different sources: 
the coupling $g$, which is 
rewritten as $g=2f\kappa^2$, and the loop integrals, from where the masses 
are combined in such a way as to produce functions of $\kappa$. 
The RG functions obtained in this way are polynomials in 
$f$ with coefficients depending on $\kappa$. Thus, it is 
not really a standard perturbative series.   
This procedure will 
be explained in Sections III and IV. 
The key point is that for 
$\kappa$ sufficiently close to $\kappa=1/\sqrt{2}$ 
from above, $f$ is small enough such that a charged fixed 
point can be obtained at $N=2$.      

An interesting point
revealed by our analysis is the role of $\kappa$ in
controlling
the appearance of
momentum space instabilities in the correlation functions.
For $\kappa>1/\sqrt{2}$,
the two-point bare correlation function
is maximized for a nonzero wave vector,
similar to the roton peak in the correlation function
of superfluid helium.
 This behavior seems to be
the
origin of the negative sign of the $\eta$-exponent and implies
a critical behavior\cite{Nogueira2}
known from the theory of Lifshitz points. In contrast
to scalar models of Lifshitz point, where the momentum space
instability is already present at the tree level,
the instability in the GL model
is induced by magnetic fluctuations.
Interestingly, such fluctuation-induced Lifshitz
points also occur in the non-commutative $\phi^4$-theory, which also
possesses a
negative $\eta$ exponent\cite{Chen}.

The plan of the paper is the following. In Section II we
discuss the advantages of  a RG approach to the GL model
in $d=3$ dimensions in the disordered phase
over the usual $4- \epsilon $-dimensional  calculations.
This approach is a generalization to the GL model
of Parisi's three-dimensional RG technique \cite{Parisi}.
In particular, we show that
this approach,
being sensitive to
the infrared divergences of the vector potential,
yields at the one-loop order a
more reliable
result than the one-loop $\epsilon$-expansion.
 In Sections III  we
renormalize the ordered phase
followed by a detailed renormalization group analysis
in Section IV which
culminates in the desired infrared-stable  fixed point
which could not be found in $4- \epsilon $ dimensions.
Some useful integrals used in the
paper compiled in the Appendix.

\section{
 Three-Dimensional
Renormalization Group in GL model
and its
Advantages
}

The bare GL hamiltonian considered in this paper is given
by

\begin{equation}
\label{GL}
{\cal H}=\frac{1}{2}({\mbox{\boldmath $\nabla$}}\times{\bf
A}_0)^2+|({\mbox{\boldmath $\nabla$}}-ie_0{\bf
A}_0)\psi_0|^2+m_0^2|\psi_0|^2+\frac{u_0}{2}|\psi_0|^4+{\cal H}_{\rm gf},
\end{equation}
where

\begin{equation}
\label{gf}
{\cal H}_{\rm gf}=\frac{1}{2\alpha_0}({\mbox{\boldmath $\nabla$}}\cdot{\bf
A}_0)^2
\end{equation}
fixes  the gauge.
It will be convenient to express the complex fields in terms
of real fields as
\begin{equation}
\psi_0=\frac{1}{\sqrt{2}}(\psi_0^{(1)}+i\psi_0^{(2)}).
\end{equation}
The propagators are given in momentum space by
\begin{equation}
G_{ij}^{(0)}({\bf p})\equiv\langle\psi_0^{(i)}({\bf p})
\psi_0^{(j)}(-{\bf p})\rangle=\frac{ \delta _{ij}}{{\bf p}^2+m_0^2},
\end{equation}

\begin{equation}
D_{\mu\nu}^{(0)}({\bf p})\equiv\langle A_0^\mu({\bf p})A_0^\nu(-{\bf p})
\rangle=\frac{1}{{\bf p}^2}\left[\delta_{\mu\nu}+(\alpha_0-1)
\frac{p_\mu p_\nu}{{\bf p}^2}\right].
\end{equation}
Since the free vector field is massless,
perturbative calculation run into infrared problems.
For instance, the
Feynman graph in Fig. 1 contributing to the four-point
function is infrared divergent at zero external momenta
for any dimension $d\in (2,4)$. Indeed,
the loop integral in this graph is proportional to
$({\bf p}^2)^{(d-4)/2}$, so that  the
Feynman integral
yields in dimensional regularization
\cite{Lawrie,Kolnberger,Folk} a pole
term  $1/\epsilon$ ($\epsilon=4-d$).
However, this pole reflects
an {\it ultraviolet} divergence
for $d=4$,
not an {\it infrared}
divergence, which is physical source of
critical properties.
Recall that the reason
why ultraviolet
divergences give nevertheless information
on the infrared behavior
is that
a
$1/\epsilon$ pole in $d=4- \epsilon $  is equivalent to a logarithmic
divergence of the form $\ln({\bf p}^2/\Lambda^2)$ at $d=4$, where
$\Lambda$ is the ultraviolet cutoff, and the ultraviolet
limit $\Lambda\to\infty$ at fixed $|{\bf p}|$
is equivalent
to taking the infrared limit $|{\bf p}|\to 0$ at fixed $\Lambda$.

Alternatively, we can  choose $|{\bf p}|\equiv \mu \neq 0$
as a scale parameter of the problem, confining the analysis to
$d\in(2,4)$ to avoid the logarithmic divergence at $d=4$.
This
procedure is applicable
in  a RG analysis {\em at\/}
the critical point
performed
by Herbut and
Tesanovic\cite{Herbut}. However, as we have discussed in the
introduction, this procedure needs a
 second scale parameter $\mu'$  which has to be determined from the outside.

Another way of regulating the infrared divergences of the GL model
is to introduce a Proca term in the Hamiltonian, $M_0^2{\bf A}_0^2/2$,
which explicitly breaks gauge invariance. Then we
may  perform the
calculations of  correlation functions using the
infrared-finite propagator
\begin{equation}
D_{\mu\nu}^{(0)}({\bf p})=
\frac{1}{{\bf p}^2+M_0^2}\left
[\delta_{\mu\nu}+(\alpha_0-1)  \frac{p_\mu p_\nu}{{\bf p}^2+\alpha_0 M_0^2
}\right],
\end{equation}
and taking $M_0\to 0$ at the end.
However, if dimensional regularization
is used to evaluate the massive Feynman integrals,
final results are the same as
in the
 dimensional regularization of the massless case, since
the $1/ \epsilon $ pole is the same for all $M_0$:
\begin{equation}
\int\frac{d^d p}{(2\pi)^d}\frac{1}{({\bf p}^2+M_0^2)^2}=
\frac{1}{8\pi^2\epsilon}+{\cal O}(1).
\end{equation}
Thus, the $\beta$-functions obtained with this procedure
will have the same
$\epsilon$-expansion as in the original
work\cite{HLM1}, where a charged fixed
point exists only
for a number of $\psi^{(i)}_0$ field components  $N>365$.

A completely different result is obtained if we
study the system directly in three dimensions.
We specify the renormalization constants
by
 rewriting the Hamiltonian (\ref{GL}) in terms of
renormalized quantities as
\begin{equation}
\label{RGL}
{\cal H}=\frac{Z_A}{2}({\mbox{\boldmath $\nabla$}}\times{\bf A})^2+
\frac{M^2}{2}{\bf A}^2+Z_\psi|({\mbox{\boldmath $\nabla$}}-ie{\bf
A})\psi|^2+Z_m m^2|\psi|^2+Z_g\frac{gm}{2}|\psi|^4+{\cal H}_{\rm gf},
\end{equation}
where the renormalized fields are defined by
$\psi=Z_\psi^{-1/2}\psi_0$ and ${\bf A}=Z_A^{-1/2}{\bf A}_0$.
We shall work
in the Landau
 gauge where $\alpha_0=0$ in ${\cal H}_{\rm gf}$. The
term $e{\bf A}$
 does not need
a renormalization
as a consequence of the
 Ward identity \cite{ZJ}
which implies that
$e^2=Z_A e_0^2$.
The renormalization constants  $Z_\psi, Z_A,Z_m$, and $Z_g$  and the
renormalized
Proca mass $M$
are fixed using the normalization
conditions for the
 one-particle irreducible
renormalized $n$-point functions\cite{ZJ,HKSF} (no sum over repeated $i$
indices):
\begin{equation}
\label{cond1}
\frac{\partial\Gamma_{ii}^{(2)}({\bf p})}{\partial{\bf p}^2}\left.\right|_
{{\bf p}=0}=1,
\end{equation}
\begin{equation}
\label{cond2}
\Gamma_{ii}^{(2)}(0)=m^2,
\end{equation}
\begin{equation}
\label{cond3}
\Gamma_{iiii}^{(4)}(0,0,0,0)=3mg,
\end{equation}
\begin{equation}
\label{cond4}
\left.\frac{\partial\Gamma_{\mu\mu}^{(2)}({\bf p})}{\partial{\bf p}^2}\right|_
{{\bf p}=0}=2,
\end{equation}
\begin{equation}
\label{cond5}
\Gamma_{\mu\mu}^{(2)}(0)=3M^2.
\end{equation}

It is straightforward to calculate the following results at the one-loop level:
\begin{equation}
\label{Zpsi}
Z_\psi=1+\frac{2}{3\pi}\frac{e_0^2}{m+M},
\end{equation}
\begin{equation}
\label{ZA}
Z_A=1-\frac{e_0^2}{24\pi m},
\end{equation}
\begin{equation}
\label{g}
g=\frac{u_0}{m}+\frac{4}{3\pi}\frac{e_0^2 u_0}{m(m+M)}
-\frac{5u_0^2}{8\pi m^2}-\frac{e_0^4}{2\pi m M}.
\end{equation}
We define a dimensionless renormalized square charge by
$f\equiv e^2/m$.
Using
$e^2=Z_A e_0^2$
we find the one-loop equation
\begin{equation}
\label{f}
f=\frac{e_0^2}{m}-\frac{e_0^4}{24\pi m^2}.
\end{equation}
The $\beta$-functions are defined
in general by  the logarithmic derivatives
\begin{equation}
\label{betafdef}
\beta_f\equiv\lim_{M\to 0}m\frac{\partial f}{\partial m},
\end{equation}

\begin{equation}
\label{betagdef}
{}~~~~~\beta_g\equiv\lim_{M\to 0}m\frac{\partial g}{\partial m}.
\end{equation}
The derivatives
with respect to $m$ have to be performed at fixed bare couplings.
At this point we make a crucial observation.
In the derivatives, we reexpress
the bare couplings in terms of the renormalization $f$ and $g$,
while discarding terms
beyond one-loop. However, the term proportional to $e_0^4$ in
Eq. (\ref{g}) contains only a single power of $m$ in the denominator.
{\it Therefore this term will contribute only to the term $-g$ in
$\beta_g$ and no term proportional to $f^2$ will be present}. The result is
\begin{eqnarray}
m\frac{\partial g}{\partial m}&=&-\frac{u_0}{m}
-\frac{4}{3\pi}\frac{e_0^2 u_0}{m(m+M)}
-\frac{4}{3\pi}\frac{e_0^2 u_0}{(m+M)^2}+\frac{5u_0^2}{4\pi m^2}
+\frac{e_0^4}{2\pi m M}\nonumber\\
&=&-\left[\frac{u_0}{m}+\frac{4}{3\pi}\frac{e_0^2 u_0}{m(m+M)}
-\frac{5u_0^2}{8\pi m^2}-\frac{e_0^4}{2\pi m M}\right]
-\frac{4}{3\pi}\frac{e_0^2 u_0}{(m+M)^2}+\frac{5u_0^2}{8\pi m^2}
\nonumber\\
&=&-g-\frac{4}{3\pi}\frac{e_0^2 u_0}{(m+M)^2}+\frac{5u_0^2}{8\pi m^2}.
\end{eqnarray}
The derivatives of $M$ with respect to $m$ do not
contribute
 since
they  are of higher than one-loop order.
Taking the physically relevant limit $M\to 0$,
 we obtain
\begin{eqnarray}
\label{betag}
\beta_g&=&-g-\frac{4}{3\pi}\frac{e_0^2 u_0}{m^2}+
\frac{5u_0^2}{8\pi m^2}\nonumber\\
&\simeq&-g-\frac{4}{3\pi}fg+\frac{5g^2}{8\pi},
\end{eqnarray}
where in the second line we have replaced $u_0/m\to g$ and
$e_0^2/m\to f$, the error committed with
this substitution is
beyond the one-loop order under consideration.
It is this result where
the three-dimensional calculation
in this paper differs essentially
from the dimensionally
regularized one in $d=4- \epsilon $ dimensions
in which
the $ \beta $-function $ \beta _g$
contains additional  $f^2$ term.
Precisely this term
 is the culprit
for the nonexistence of a charged
fixed point in HLM.

The second  $\beta$-function is
\begin{equation}
\label{betaf}
\beta_f=-f+\frac{f^2}{24\pi},
\end{equation}
{}From Eqs. (\ref{betag}) and (\ref{betaf})
we obtain the infrared stable fixed
point $f_*=24\pi$ and $g_*=264\pi/5$.

At this point we may think that
the problem is solved,
 and that the absence
of a charged fixed point was  merely  an artifact of the
$\epsilon$-expansion.
Unfortunately this is not true since at the fixed point,
{\it the critical exponents have completely unphysical values}.
For example, the critical exponent $\eta$, which is defined
by the fixed point value of the RG function
\begin{equation}
\gamma_\psi\equiv\lim_{M\to 0}m\frac{\partial\ln Z_\psi}{\partial m},
\end{equation}
has the value $\eta=-16$, which is unphysical since
it does not respect the bound $\eta>-1$.
The reason for this
is that the fixed-point value $f_*$ is
quite
large, and this leads to the unphysical value
of $  \eta $.

Does this mean that the
present approach must be discarded as well?
Not completely. We are still
much better off than HLM.
As mentioned before,
they judged
 how  far their result
was from nature,
by changing the number
of complex  fields from 1 to $N/2$
with an O($N$) symmetry
among the real components.
This served  to suppress
the $f^2$ term, since the graph of Fig. 1 is
of order\cite{Note1} $1/N^2$. Our calculation
has the advantage that
 the $f^2$ term is absent
 for any $N$.
Suppose we go through our analysis
for  $N>2$. Then we find
\begin{equation}
\label{betafN}
\beta_f=-f+\frac{Nf^2}{48\pi},
\end{equation}
\begin{equation}
\label{betagN}
\beta_g=-g-\frac{4fg}{3\pi}+\frac{(N+8)g^2}{16\pi}.
\end{equation}
The critical exponent has now the value
 $\eta=-32/N$, which is physically meaningful provided
$N>N_c=32$.
Thus our three-dimensional approach yields a critical $N$
value $N_c=32$,
where the result becomes physical. This lies much lower than the critical
HLM one-loop value \cite{HLM1} $N_c^{\rm HLM}=365$.
If our
one-loop value $N_c=32$ is lowered
by the two-loop corrections,
there is a good chance of getting
a physical exponent $ \eta $
for a single complex order field $\psi$.
Such a calculation remains to be done.

Further insight can be obtained by calculating the critical exponent $\nu$. 
This is given by 

\begin{equation}
\frac{1}{\nu}=2+\gamma_m^*-\eta,
\end{equation}
where $\gamma_m^*$ is the fixed point value of the RG function 

\begin{equation}
\gamma_m\equiv m\frac{\partial\ln Z_m}{\partial m}.
\end{equation}
At one-loop order it is given by

\begin{equation}
\gamma_m=-\frac{(N+2)g}{16\pi}.
\end{equation}
Thus, 

\begin{equation}
\frac{1}{\nu}=2-\frac{(N+2)(N+64)}{N(N+8)}+\frac{16}{N}.
\end{equation}
Positive values of $\nu$ are obtained for $N>34$. However, 
for any finite value of $N$ we have $\nu>1$. When $N\to\infty$ 
we have $\nu=1$.

As a last remark in this section, let us comment about the
absence of a tricritical fixed point in our one-loop calculation.
The tricritical fixed point is absent because there is no
$f^2$ in the $\beta$-functions (\ref{betag}) and (\ref{betagN}).
With this respect, our approach gives a result similar to the
$1/N$ expansion at ${\cal O}(1/N)$, where there is only a second-order
fixed point and no tricritical fixed point. It is evident, however,
that a tricritical fixed point will be generated at two-loops, but
in this case resummation is necessary. The same result is to be
expected in the $1/N$ expansion at ${\cal O}(1/N^2)$.

\section{Renormalization constants in the ordered phase
}

\subsection{Problems with unitary gauge}

We have seen in the preceding section that
in three dimensions the situation improves
by using a massive vector field to
 avoid
 infrared divergences in the Feynman integrals.
Above $T_c$, this mass is set equal to zero at the end.
In the ordered phase, such a mass
is automatically present
as a consequence of the Meissner effect.
In fact,
there are no massless modes at all in the ordered phase,
the Goldstone boson supplying
the
 longitudinal component to the massive vector field.
The simplest way of
representing this effect is in the
unitary gauge.
At the mean-field level,
this  amounts to the field
parametrizations (dropping the subscripts 0 indicating
bare quantities for simplicity)
\begin{equation}
\psi=\frac{1}{\sqrt{2}}(v+\rho)e^{i\theta}, \label{@rho}
\end{equation}
\begin{equation}
\label{gtransf}
{\bf B}={\bf A}+\frac{1}{e}{\mbox{\boldmath $\nabla$}}\theta.
\end{equation}
The above parametrization allows for a nonzero expectation
value of the order field, that is, by setting
$\langle\psi\rangle=v$. At the tree level this happens for
$m^2<0$. To avoid proliferating minus signs,
it will be more convenient
in this section
to replace the term $m^2|\psi|^2$ in the Hamiltonian by
$-m^2|\psi|^2$, such that a nonzero expectation value
of the order field exists for $m^2>0$.
In the Coulomb gauge ${\mbox{\boldmath $\nabla$}}\cdot{\bf A}=0$
the partition function is calculated from
the following functional integral
\begin{equation}
Z=\int{\cal D}{\bf A}
{\cal D}\rho\, {\cal D}\theta \rho\,
\det(-\partial^2)\delta({\mbox{\boldmath $\nabla$}}\cdot{\bf A})
 \,e^{-\int d^3 r{\cal H}},
\label{@}\end{equation}
the last factor  in the measure
being the
 Faddeev-Popov determinant.
This is  a convenient formulation of the
field system since by performing the change of
variables (\ref{@rho}) and (\ref{gtransf})
the functional integral can immediately be simplified
by integrating out the angular field $\theta$. Its value is
fixed by the delta function enforcing the Coulomb gauge.
The result of this integral  precisely
cancel  the
 Faddeev-Popov determinant.

In terms of the fields
(\ref{@rho}), the Hamiltonian becomes
\begin{equation}
\label{unitH}
{\cal H}=V_0(v)+\frac{1}{2}({\mbox{\boldmath $\nabla$}}\times{\bf B})^2
+\frac{m_A^2}{2}{\bf B}^2+\frac{1}{2}({\mbox{\boldmath $\nabla$}}\rho)^2
+\frac{e^2}{2}\rho^2{\bf B}^2+e^2v\rho{\bf B}^2+
m^2\rho^2+\frac{uv}{2}\rho^3+\frac{u}{8}
\rho^4,
\end{equation}
where $V_0(v)=-m^2v^2/2+uv^4/8$, $m_A^2=e^2v^2$, and
$m^2=uv^2/2$. In this well known tree level analysis
 the Goldstone has disappeared and the
vector potential has becomes massive.

The
free propagator of the  vector field
  is
given by
\begin{equation}
D^{(0)}_{\mu\nu}({\bf p})=\frac{1}{{\bf p}^2+m_A^2}\left(\delta_{\mu\nu}
+\frac{p_\mu p_\nu}{m_A^2}\right),
\end{equation}
which corresponds to take the gauge-fixing parameter
$\alpha\to\infty$.
Now, at large $|{\bf p}|$ the above propagator behaves like a constant. Thus,
the free scaling dimension of ${\bf B}$ does not have the
typical canonical
 scaling dimension  $1/2$ of a vector field, but
$3/2$.
But then
power counting tells us that the interaction
$e^2\rho^2{\bf B}^2/2$ is not renormalizable in $d=3$.
Thus, in the unitary gauge, the
absorption
of the Goldstone mode in the longitudinal part of the
massive vector meson seems to
destroy renormalizability. The only way to obtain a renormalizable
theory out of the unitary gauge is by keeping the gauge-fixing
parameter $\alpha$ nonzero during the Feynman integrals calculation
and take the limit $\alpha\to\infty$ at the end \cite{Collins}.
In order to implement such a program, we should not use the
unitary gauge parametrization given by Eqs. (\ref{@rho}) and
(\ref{gtransf}). This means that the Goldstone boson will still be
present in the theory. The pole coming from the Goldstone boson
propagator will be, however, canceled from physical quantities.

Beside the renormalizability problem,
there is another difficulty with the unitary gauge which
is related to the vortex content of the theory.
The point is that the covariant derivative term
$|({\mbox{\boldmath $\nabla$}}-ie{\bf
A})\psi|^2$
 in the GL Hamiltonian
(\ref{GL})
goes over, in the parametrizations (\ref{@rho}) and (\ref{gtransf}),
into
$|({\mbox{\boldmath $\nabla$}}-ie{\bf
B}-2\pi
{\mbox{\boldmath $\theta$}_v}
) \rho|^2$,
where $
{\mbox{\boldmath $\theta$}}_v
$  is
 the {\em vortex gauge field\/},
a vector field describing the vortices in the superfluid
\cite{KleinertBook,Cam}.
This field arises from the fact that
 $\theta$ is not a single-valued field,
but a
a cyclic field with the property
 $\int d\theta({\bf r})=2\pi n_v({\bf r})$,
where $n_v$ is the winding number or
vorticity.
For this reason,
the gradient $\nablab e^{i\theta}$
is not simply
equal to
 $i(\nablab \theta) e^{i\theta}$,
   but
 must be supplemented
by such a vortex gauge field yielding
$i(\nablab \theta-2\pi\thetab_v ) e^{i\theta}$,
and ensuring the periodicity under $\theta\rightarrow \theta+2\pi$.
The vector gauge field
 removes possible delta functions
arising from the gradient of jumps by $2\pi$ in the cyclic field
which exist around a vortex line
in the superconducting phase  \cite{KleinertBook,Cam}.
In fact,  these vortices  are crucial for driving the phase transition
and can therefore not be neglected.
It is, however, difficult to treat the vortex gauge field
by
conventional perturbation theory
using
Feynman diagrammatic techniques.
If we want to take vortices into account
in  perturbation expansions
we must therefore avoid multivalued fields.

\subsection{The Landau gauge}

A  field
 parametrization which avoids the
problems of the
unitary gauge and has
more desirable  properties
with respect to power counting
is
\begin{equation}
\psi_0=\frac{1}{\sqrt{2}}(v_0+\sigma_0+i\pi_0).
\end{equation}
Here the bare Hamiltonian becomes
\begin{eqnarray}
\label{orderedGL}
{\cal H}&=&{\cal H}_{\rm free}+e_0{\bf A}_0\cdot(\sigma_0{\mbox{\boldmath
$\nabla$}}\pi_0-\pi_0
{\mbox{\boldmath $\nabla$}}\sigma_0)
+e_0^2v_0\sigma_0{\bf A}_0^2+\frac{e_0^2}{2}(\sigma_0^2+\pi_0^2){\bf A}
_0^2\nonumber\\
&+&\frac{u_0v_0}{2}\sigma_0^3+\frac{u_0v_0}{2}\sigma_0\pi_0^2+
\frac{u_0}{8}(\sigma_0^2+\pi_0^2)^2,
\end{eqnarray}
where ${\cal H}_{\rm free}$
denotes the free part of the
Hamiltonian:

\begin{eqnarray}
\label{quadH}
{\cal H}_{\rm free}&=&\frac{1}{2}({\mbox{\boldmath $\nabla$}}\times{\bf
A}_0)^2+\frac{m_{A,0}^2}{2}{\bf A}_0^2
+\frac{1}{2}[({\mbox{\boldmath $\nabla$}}\sigma_0)^2+({\mbox{\boldmath
$\nabla$}}\pi_0)^2]+
\frac{1}{2}(-\bar{m}_0^2+3m_0^2)\sigma_0^2\nonumber\\
&+&\frac{1}{2}(-\bar{m}_0^2+m_0^2)
\pi_0^2+J_0\sigma_0+{\cal H}_{\rm gf}.
\end{eqnarray}
The Coulomb gauge ${\mbox{\boldmath $\nabla$}}\cdot{\bf A}_0=0$
is fixed by letting $\alpha_0\to 0$, which corresponds
to the so called Landau gauge.
We have therefore
 omitted
 a crossed term $e_0{\bf A}_0\cdot{\mbox{\boldmath $\nabla$}}\pi_0$.
The bare masses are $m_0^2=u_0v_0^2/2$ and $m_{A,0}^2=e_0^2v_0^2$.
We have introduced a source term for the longitudinal field
$\sigma_0$. In the equation of motion, a zero
source corresponds to the minimum of the effective action. At the
tree level we have $J_0=v_0(-\bar{m}_0^2+m_0^2)$. Thus, the
tree level minimum implies $\bar{m}_0^2=m_0^2$.

{}From ${\cal H}_{\rm free}$ we obtain the propagators:

\begin{equation}
\label{HiggsProp}
G_{\sigma\sigma}^{(0)}({\bf p})=\frac{1}{{\bf p}^2+3m_0^2-\bar{m}_0^2},
\end{equation}

\begin{equation}
\label{GoldProp}
{}~~~~~G_{\pi\pi}^{(0)}({\bf p})=\frac{1}{{\bf p}^2+m_0^2-\bar{m}_0^2},
\end{equation}

\begin{equation}
\label{PhotProp}
D_{\mu\nu}^{(0)}({\bf p})=\frac{1}{{\bf p}^2+m_{A,0}^2}\left(
\delta_{\mu\nu}-\frac{p_\mu p_\nu}{{\bf p}^2}\right).
\end{equation}
Now power counting gives the desired
dimensions in the ultraviolet  since
the vector propagator behaves like $\sim 1/{\bf p}^2$
for large $|{\bf p}|$. Moreover, the vector field
is
massive and the graph of Fig. 1 is convergent in $d=3$.
However, we still have a massless mode, the $\pi$-field.
Interestingly, in the one-loop calculation
of the vector potential and $\pi$ two-point functions,
$\Gamma_{\mu\nu}^{(2)}$ and $\Gamma_{\pi\pi}^{(2)}$,
respectively, the infrared divergences from the would-be Goldstone
boson do not play any role.  We shall
calculate here the bare two-point functions. The Feynman
graphs contributing to these two-point functions are shown in
Fig. 2. The only ultraviolet divergences come from the tadpoles and
they are all proportional to the ultraviolet cutoff $\Lambda$. These
divergences are absorbed into the bare masses $m_0^2$ and
$m_{A,0}^2$. Also, in the loop integrals we shall replace
the bare masses by the renormalized ones, since the error
involved in this replacement is of higher order.

The renormalized two-point functions
are given in terms of the bare ones by
$\Gamma_{\sigma\sigma}^{(2)}=Z_\sigma\Gamma_{0,\sigma\sigma}^{(2)}$,
$\Gamma_{\pi\pi}^{(2)}=Z_{\pi}\Gamma_{0,\pi\pi}^{(2)}$,
and $\Gamma_{\mu\nu}^{(2)}=Z_{A}\Gamma_{0,\mu\nu}^{(2)}$.
In order to determine the renormalization
constants $Z_{\pi}$ and $Z_A$ we employ the
normalization conditions

\begin{equation}
\label{condpi2pZ}
\left.\frac{\partial\Gamma_{\pi\pi}^{(2)}({\bf p})}{\partial{\bf p}^2}
\right|_{{\bf p}=0}=1,
\end{equation}

\begin{equation}
\label{condA2pZ}
\left.\frac{\partial\Gamma_{\mu\mu}^{(2)}({\bf p})}{\partial{\bf p}^2}
\right|_{{\bf p}=0}=2.
\end{equation}
Since $\pi$ is a would-be Goldstone boson, we require also the
normalization conditions for renormalized source $J=Z_\sigma^{1/2}J_0=0$:

\begin{equation}
\label{condsig2p}
\Gamma_{\sigma\sigma}^{(2)}(0)=uv^2\equiv2m^2,
\end{equation}

\begin{equation}
\label{condpi2p}
\Gamma_{\pi\pi}^{(2)}(0)=0,
\end{equation}
such that all
 $\pi$-propagators
in loop integrals are
massless.

The following normalization condition defines the renormalized
mass of the vector field
\begin{equation}
\label{condA2p}
\Gamma_{\mu\mu}^{(2)}(0)=3e^2v^2\equiv 3m_A^2.
\end{equation}
Note that, as usual, the renormalization conditions are chosen
in such a way as to be consistent
with the tree-level and the Ward identities \cite{Lee}.
Note that the above
conditions determine completely the renormalization constants. Since we
have five renormalization constants, the above five conditions determine
them completely. Actually, the Ward identities imply that
$Z_\sigma=Z_\pi$ if the Coulomb gauge is chosen, a result consistent
with the gauge invariance of the Hamiltonian.
The renormalization
conditions imply that the renormalized Ginzburg parameter
is given by $\kappa^2=m^2/m_A^2=u/2e^2$, that is, it has the same form
as at the tree-level but with the bare couplings replaced by the
renormalized ones.

After some work we obtain
\begin{eqnarray}
\label{2pfpi}
\Gamma_{0;\pi\pi}^{(2)}({\bf p})&=&{\bf p}^2-\bar{m}_0^2+m_0^2
-\frac{u_0\sqrt{2}m}{8\pi}-\frac{e_0^2 m_A}{2\pi}
\nonumber
\\*[0.8cm]
&-&\frac{u_0 m^2}{4\pi|{\bf p}|}b({\bf p},\sqrt{2}m)
-e_0^2\left\{
\frac{1}{8\pi m_A^2|{\bf p}|}[4m_A^2{\bf p}^2
+({\bf p}^2+2m^2-m_A^2)^2]
a({\bf p},\sqrt{2}m,m_A)\right.\nonumber
\\*[0.8cm]
&+&\left.\frac{{\bf p}^2}{4\pi m_A}-\frac{1}{8\pi m_A^2|{\bf p}|}(
{\bf p}^2+2m^2)^2 b({\bf p},\sqrt{2}m)\right\},
\end{eqnarray}

\vspace{0.5cm}

\begin{eqnarray}
\label{2pfA}
\Gamma_{0;\mu\nu}^{(2)}({\bf p})&=&\delta_{\mu\nu}\left\{{\bf p}^2+m_{A,0}^2-
\frac{e_0^2 m_A}{8\pi}(\sqrt{2}\kappa-1)-\frac{e_0^2 m_A^3}{8\pi{\bf p}^2}
(1-2\kappa^2-\sqrt{2}\kappa)-\frac{e_0^2\sqrt{2}m^3}{4\pi{\bf p}^2}
\right.\nonumber
\\*[0.8cm]
&+&\left.\frac{e_0^2}{16\pi}\frac{[{\bf p}^2+m_A^2(\sqrt{2}\kappa+1)^2]
[{\bf p}^2+m_A^2(\sqrt{2}\kappa-1)^2]-8m_A^2{\bf p}^2}{|{\bf p}|^3}
a({\bf p},\sqrt{2}m,m_A)\right\}
\nonumber
\\*[0.8cm]
&+&\frac{p_\mu p_\nu}{{\bf p}^2}\left\{-{\bf p}^2+\frac{e_0^2 m_A}{8\pi}(
\sqrt{2}\kappa-3)+\frac{3e_0^2\sqrt{2}m^3}{4\pi{\bf p}^2}
+\frac{3e_0^2m_A^3}{8\pi{\bf p}^2}(1-2\kappa^2-\sqrt{2}\kappa)
\right.\nonumber
\\*[0.8cm]
&-&\frac{e_0^2 m^4}{2\pi|{\bf p}|^3}b({\bf p},\sqrt{2}m)
-\frac{e_0^2}{16\pi|{\bf p}|^3}[3({\bf p}^2+2m^2)^2-12m^2m_A^2
\nonumber
\\*[0.8cm]
&+&\left.3m_A^4
-2m_A^2{\bf p}^2]a({\bf p},\sqrt{2}m,m_A)\right\}
\end{eqnarray}
with the functions
\begin{eqnarray}
\label{a}
a({\bf p},m_1,m_2)&=&
\arctan\left(\frac{{\bf p}^2+m_1^2-m_2^2}{2m_2|{\bf p}|}
\right)+\arctan\left(\frac{{\bf p}^2+m_2^2-m_1^2}{2m_1|{\bf p}|}
\right),
\\
\label{b}
b({\bf p},m_1)&=&\lim_{m_2\to 0}a({\bf p},m_1,m_2)\nonumber\\
&=&\frac{\pi}{2}+\arctan\left(\frac{{\bf p}^2-m_1^2}{2m_1|{\bf p}|}\right),
\end{eqnarray}
where $\kappa\equiv m/m_A=\lambda/\xi$ is the renormalized Ginzburg parameter.
The bare Ginzburg parameter is given by $\kappa_0= \sqrt{u_0/2e_0^2}$,
since $m_0^2=u_0 v_0^2/2$ and $m_{0,A}^2=e_0^2v_0^2$.

On account of the normalization condition
(\ref{condpi2p}), we obtain the one-loop correction to $\bar{m}_0^2$:
\begin{equation}
\label{mbar2}
\bar{m}^2=m_0^2-\frac{3u_0\sqrt{2}m}{8\pi}-\frac{e_0^2 m_A}{2\pi}.
\end{equation}

In the following, we  shall use extensively the replacement
$\kappa_0^2\to\kappa^2=m^2/m_A^2=u^2/2e^2=g/2f$, where
$f=e^2/m$ and $g=u/m$ are dimensionless renormalized couplings,
neglecting
all errors of higher  than one-loop order.
The parameter  $\kappa$ arises
 in two different contexts:
once from the renormalized mass ratio $m/m_A$,
and once from the ratio of couplings as $ \sqrt{u/2e^2}$.

We shall then parametrize all  RG functions in terms
of $\kappa$ and $f$ instead of $g$ and $f$.
Our expansion will be controlled initially by powers of
$f$, but once $g$ is eliminated
in favor of $\kappa$,
then $\kappa$
 will no longer be assumed to be small, such that
the RG functions
are
not
polynomials
in $f$ and $\kappa$ as
in the previous expansions
in powers
of  $f$
and $g$ in Section II.

Below we shall see that in this formulation
there exists a natural expansion parameter related to $\kappa$,
namely $ \Delta  \kappa \equiv  \kappa -1/ \sqrt{2}$.
Remarkably,
the present  description of the fixed-point structure
requires
only the
knowledge of the two-point functions, since
from Eq. (\ref{mbar2}) $m^2=Z_\pi\bar{m}^2$ and
$e^2=Z_A e_0^2$, with $Z_A$ being determined from the renormalization
condition (\ref{condA2pZ}) (see below). This
represents an immense advantage
since the four-point
functions contain severe infrared singularities
coming from the
would-be Goldstone boson. Of course, we expect  these
singularities to cancel at the end,
 but this is a
difficult issue in fixed dimensions. In the
$4- \epsilon $ -dimensional regularization scheme
this issue is avoided
since
 the counterterms required for a renormalization
in
the
ordered phase are identical to those
in the disordered
phase where they  can
be calculated without Goldstone-bosons \cite{ZJ}.

{}From the normalization condition (\ref{condpi2pZ}) we
derive
\begin{equation}
\label{Zperp}
Z_\pi=1-\frac{f \kappa(2\kappa ^2+ \sqrt{2}\kappa -8)}
{12\pi(\sqrt{2}\kappa+1)},
\end{equation}
which satisfies $Z_\pi>1$  for
$\kappa< \hat \kappa \equiv (\sqrt{33}-1)/2\sqrt{2}\approx1.6774$.
Only for $\kappa> \hat \kappa $ will
 the
wave function renormalization satisfy the usual
inequality $0\leq Z_\pi<1$
 found in the absence of  gauge fields.

Let us now calculate $m_A^2$ at one-loop level using
the normalization condition (\ref{condA2p}). We find

\begin{eqnarray}
\frac{1}{3}\Gamma_{0,\mu\mu}^{(2)}&=&m_{A,0}^2+
\frac{e_0^2m_A}{12\pi(\sqrt{2}\kappa+1)}(2\kappa^2+\sqrt{2}\kappa-8)
\nonumber\\
&\simeq&Z_{\pi}^{-1}m_{A,0}^2.
\end{eqnarray}
Since $\Gamma_{\mu\nu}^{(2)}=Z_A\Gamma_{0,\mu\nu}^{(2)}$, we
obtain

\begin{equation}
\label{mA2}
m_A^2=Z_A Z_\pi^{-1}m_{A,0}^2.
\end{equation}
Thus $Z_\pi$ appears in the renormalization
of the mass $m_A$
of the vector field. This result is expected from
the Ward identities. It provides us with
 a good check of the
consistency of our calculations. Physically it shows clearly
how the
fluctuations of the
Goldstone boson renormalize the
mass of the
vector field, which has been exploited
in the theory of electroweak interactions
to build a renormalizable theory
of massive vector mesons and Higgs fields. Thus, although the
Goldstone boson is present in the calculations, the Higgs
mechanism takes care of absorbing its
fluctuating degrees of freedom to build a fluctuating longitudinal
degree of freedom for the vector potential.

The superfluid density is intimately related to
the penetration depth $\lambda=1/m_A$. It is
given by

\begin{equation}
\label{superfdens}
\rho_s=Z_\pi^{-1}v_0^2,
\end{equation}
which reflects a relation due to Josephson\cite{Josephson}.
On account of
(\ref{mA2})
and
the relation
$e^2=Z_A e_0^2$
 we can write $m_A^2=e^2\rho_s$.

It remains to
calculate $Z_A$. This is done using the normalization
condition (\ref{condA2pZ}), yielding
\begin{equation}
\label{ZAord}
Z_A=1-\frac{\sqrt{2}C(\kappa)f}{24\pi(2\kappa^2-1)^3},
\end{equation}
where

\begin{equation}
C(\kappa)=4\kappa^6+10\kappa^4-24\sqrt{2}\kappa^3
+27\kappa^2+4\sqrt{2}\kappa-{1}/{2}.
\end{equation}
We note that the second term in Eq. (\ref{ZAord}) is singular
at $\kappa=1/\sqrt{2}$. This singularity is analogous to the
$1/\epsilon$ singularity in dimensionally regularized theories.
The important role of this singularity will be discussed in more detail
in the next Section.

\section{Renormalization group analysis in the ordered phase. Role
of the Ginzburg parameter $\kappa$}

\subsection{Renormalization group functions}

We are now prepared to calculate the RG functions. Let us define the
RG functions:

\begin{equation}
\label{gammapi}
\gamma_\pi\equiv m\frac{\partial\ln Z_\pi}{\partial m},
\end{equation}

\begin{equation}
\label{gammaA}
\gamma_A\equiv m\frac{\partial\ln Z_A}{\partial m}.
\end{equation}
The RG functions $\gamma_\pi$ and $\gamma_A$ are given
explicitly
at one-loop order as

\begin{equation}
\label{gammapi1loop}
\gamma_\pi=\frac{\kappa\,f}{12\pi}\frac{2 \kappa ^2+ \sqrt{2} \kappa -8
}{ \sqrt{2} \kappa +1},
\end{equation}

\begin{equation}
\label{gammaA1loop}
\gamma_A=\frac{\sqrt{2}C(\kappa)f}{24\pi(2\kappa^2-1)^3}.
\end{equation}
Note that while deriving Eqs. (\ref{gammapi1loop}) and
(\ref{gammaA1loop}) we have not differentiated $\kappa$, since
this would lead
to higher powers of $f$  beyond the one-loop approximation.
We observe also that the singularity at $\kappa=1/\sqrt{2}$ present
at $Z_A$ is also present in $\gamma_A$.

The $\beta$-function of $f$ is then given by

\begin{equation}
\label{betaforder}
\beta_f=(\gamma_A-1)f.
\end{equation}
To obtain the $\beta$-function of $\kappa^2$ we need to know the
evolution equation for $m_A^2$. From Eq.~(\ref{mA2}) we obtain

\begin{equation}
\label{flowmA0}
m\frac{\partial m_A^2}{\partial m}=(\gamma_A-\gamma_\pi)m_A^2+
Z_AZ_\pi^{-1}m\frac{\partial m_{A,0}^2}{\partial m},
\end{equation}
thus reducing the problem to a calculation of
$m\partial m_{A,0}^2/\partial m$. It must be emphasized that
in our approach this is not zero.
Our differentiations are performed at fixed $ \kappa _0^2=m_0^2/m_{A,0}^2$,
such that
$m\partial\kappa_0^2/\partial m=0$
whereas  both $m_0$ and $m_{A,0}$ remain functions of $m$.
Due to the fixed $ \kappa _0$, their derivatives
$m\partial m_{A,0}^2/\partial m=\zeta m_{A,0}^2$ and
$m\partial m_{0}^2/\partial m=\zeta m_0^2$,
are governed by a common RG function
 $\zeta$.
 In order to obtain
$m\partial m_{0}^2/\partial m$ we use Eq. (\ref{mbar2}) to
write

\begin{equation}
\label{m02}
m_0^2=Z_\pi^{-1}m^2+\frac{3u_0\sqrt{2}m}{8\pi}+\frac{e_0^2 m_A}{2\pi},
\end{equation}
where we have inserted $m^2=Z_\pi\bar{m}^2$.
The derivative yields

\begin{equation}
\label{flowm02}
m\frac{\partial m_0^2}{\partial m}=(2+\zeta_\pi-\gamma_\pi)m_0^2,
\end{equation}
where up to one loop

\begin{equation}
\label{gammam}
\zeta_\pi
=-\frac{\sqrt{2}}{4\pi}f\left(\frac{3\kappa^2}{2}+
\frac{1}{\sqrt{2}\kappa}\right).
\end{equation}
{}From Eqs. (\ref{flowmA0}) and (\ref{flowm02}) we derive

\begin{equation}
\label{flowmA}
m\frac{\partial m_A^2}{\partial m}=(2+\zeta_\pi+\gamma_A-2\gamma_\pi)m_A^2.
\end{equation}

It is  now straightforward to obtain the $\beta$-function for
$\kappa^2$:

\begin{equation}
\label{betak2}
\beta_{\kappa^2}\equiv m \frac{\partial  \kappa^2 }{\partial m}= m
 \frac{\partial  m^2/m_A^2 }{\partial m}= \kappa ^2\left(2-
\frac{m}{m_A^2}\frac{\partial m_A^2}{\partial m}\right)=
\kappa^2(2\gamma_\pi-\gamma_A-\zeta_\pi).
\end{equation}

{}From Eqs. (\ref{betaforder}) and (\ref{betak2}) we see that
charged fixed points satisfy the equations
\begin{equation}
\eta_A\equiv \gamma_A(f_*,\kappa_*)=1
 \label{@fp1}\end{equation}
and
\begin{equation}
2\eta-1=\zeta_\pi(f_*,\kappa_*),  ~~~{\rm where}~~~
\eta\equiv\gamma_\pi(f_*,\kappa_*).
 \label{@fp2}\end{equation}
 The exponents
$\eta_A$ and $\eta$ determine the anomalous scaling dimensions
of the fields ${\bf A}$ and $\pi$, respectively.
Our equations yield  fixed points at
\begin{equation}
f_*\approx 0.3, ~~~~~~\kappa_*\approx 1.17/\sqrt{2}.
\label{@fpvs}\end{equation}
At this fixed point the value of the $\eta$-exponent is
$\eta\approx-0.02$,
which fulfills the inequality $\eta>-1$, showing that our
fixed point in the ordered phase is completely physical,
in contrast to the $N=2$ case in the disordered phase
calculation of Section II.

Our $\eta$-exponent is less negative than most values found
in the literature. There is, however, no consensus about this value.
For instance, a recent Monte Carlo
simulation \cite{Hove} give $\eta\approx -0.24$, disagreeing with
another  Monte Carlo value
 $\eta\approx -0.79$ published recently\cite{Olsson1}, which is
similar to $\eta=-0.74$ found by
Herbut and Tesanovic by adjusting
their parameter $c$
 to fit the Kleinert's value of the tricritical Ginzburg parameter,
$\kappa_t=0.77/\sqrt{2}$. The value $\eta=-0.2$ is obtained
when the $c$-parameter is fitted to
the now less precise (see
Ref. \onlinecite{SUD}) Monte Carlo value \cite{Barth}
$\kappa_t=0.42/\sqrt{2}$.
There are other values of $\eta$ reported in
the literature \cite{Radz,Berg}, exhibiting
a lack of
concensus on its numerical value.

\subsection{Validity of the method}

We now want to explain what makes
our three-dimensional procedure  so successful in
finding the charged fixed points.
The main obstacle in our calculation in the disordered
phase in Section II was
the too large value of $f_*$.
In previous approaches in $4- \epsilon $ dimensions,
this prevented a fixed point to exist.
The size of $f^*$  was diminished
by a factor $2/N$
via the artifical extension to $N/2$ complex fields, and this led to a fixed
point after all for  $N>N_c=365$.
The three-dimensional calculation in the disordered phase
is far less unphysical
since it always give a fixed point.
Still, there was the
problem that the associated critical exponent $ \eta $ was too negative,
violating the bound $\eta>-1$,
again because of a too large $f^*$,
unless $N>N_c=32$.

These difficulties are absent
in three dimensions below $T_c$.
To understand this let us define an
{\it effective square charge} $\bar{f}(\kappa)$ depending on $ \kappa $ by the
equation

\begin{equation}
\label{fbareq}
\gamma_A(\bar{f},\kappa)=1,
\end{equation}
meaning that we go to the ``fixed line''
in the two-dimensional space of coupling constants defined by the vanishing of
the $ \beta $-function $ \beta _f$ in Eq. (\ref{betaforder}),
but not at the fixed point for Eq. (\ref{betak2}). The $ \kappa $-dependence of
$\bar{f}(\kappa)$  is

\begin{equation}
\label{fbar}
\bar{f}(\kappa)=\frac{24\pi(2\kappa^2-1)^3}{\sqrt{2}C(\kappa)}.
\end{equation}
We see that $\bar{f}(\kappa)=0$ for $\kappa=1/\sqrt{2}$, which is
just the value separating the type I
from the type II regime  at the mean-field level.
The $ \kappa $-dependence of $ \bar f( \kappa )$
is plotted in Fig.~\ref{fbarfig}.
The effective square charge is negative if
 $0.096/\sqrt{2}<\kappa<1/\sqrt{2}$ and
very large positive for
$0\leq  \kappa<0.096/\sqrt{2}$. This
 makes it
 {\em impossible\/} to have an infrared-stable
positive charge in
  the type-I regime
 $\kappa<1/\sqrt{2}$,
and we conclude that
the
phase
transition is of first-order in the type-I regime.
Moreover, we observe
that for  $\kappa$ slightly above
$1/\sqrt{2}$, the
effective critical charge is small.

An important remark is in order here.
In Fig. \ref{fbarfig} there is an asymptote which separates
two distinct regions: one where the charged fixed point is
inaccessible (the left side of the asymptote), and another
one where the charge fixed point is accessible. This
asymptote reveals a Landau-ghost-like behavior. The
so called ``runaway'' of the RG flow\cite{HLM1} corresponds
to the left side of the asymptote. It is worth mentioning
that a similar feature can be expected in the
$\beta$-function of QCD\cite{Grunberg}.
There the asymptote is supposed to separate the asymptotically free
regime from the regime containing  an infrared stable fixed
point which governs quark confinement.

Consider now the large-$\kappa$ limit
which corresponds to an extreme type II regime. There are two
ways of taking $\kappa\to\infty$: either by letting $g\to\infty$ at
$f$ fixed, or by letting $f\to 0$ at $g$ fixed. In the first
case we obtain from
Eq. (\ref{gammaA}) $\gamma_A|_{\kappa=\infty,f}=f/(24\pi\sqrt{2})$,
while in the second: $\gamma_A|_{\kappa=\infty,g}=0$.
In the first case the magnetic fluctuations are still relevant
and the $|\psi|^4$ interaction may be replaced by a
pure phase field with fixed
absolute value of the order parameter, i.e.,
 the GL model be be approximated by its London limit.
In the second case the magnetic
fluctuations disappear completely
from the system and the universality
class is that of a pure $|\psi|^4$-theory which is equal
to that of the $XY$-model.

By substituting  $\bar{f}( \kappa )$ of Eq.~(\ref{fbar}) into
 (\ref{gammapi1loop}),
we obtain the effective exponent $\bar{\eta}(\kappa)$.
For $\kappa=1/\sqrt{2}$,
this has the mean-field value $ \eta =\bar{\eta}(1/ \sqrt{2}) =0$.
In the physically interesting interval
 $1/\sqrt{2}<\kappa<\hat  \kappa \equiv (\sqrt{33}-1)/2\sqrt{2}
\approx1.67746$,
the effective exponent $\bar{\eta}( \kappa )$ is {\em negative\/}, as shown
in
Fig.~\ref{f2}. Above $ \hat  \kappa $, it is positive.
Remarkably, it
 lies above   the physical lower bound
$-1$
for all $\kappa>1/ \sqrt{2} $, i.e., in the entire
type-II regime in the mean-field approximation.

Thus,  our
RG approach in the ordered phase gives
effectively  an expansion around $\kappa=1/\sqrt{2}$. It is therefore
convenient
to
introduce the expansion parameter
$\Delta\kappa\equiv\kappa-1/\sqrt{2}$. Then we
can write the leading expansion term in $\Delta\kappa$ for
the effective square charge $\bar{f}(\kappa)$ and
the effective exponent $\bar{\eta}(\kappa)$ as

\begin{equation}
\label{fexpansion}
\bar{f}(\kappa)=48\pi\Delta\kappa^3,
\end{equation}

\begin{equation}
\label{etaexpansion}
\bar{\eta}(\kappa)=-6\sqrt{2}\Delta\kappa^3.
\end{equation}
Similarly, we expand

\begin{equation}
\label{betak2expansion}
\beta_{\kappa^2}(\kappa^2,\bar{f}(\kappa))=
-\frac{1}{2}-\sqrt{2}\Delta\kappa-\Delta\kappa^2+162\sqrt{2}
\Delta\kappa^3.
\end{equation}

We now discuss
the other important critical exponent
$ \nu $.
Recall that it is defined in general
by the scaling relation
\begin{equation}
 \nu =\frac{1}{2- \eta _m},
\label{@nu}\end{equation}
where  $ \eta _m$ is the fixed-point value of the logarithmic derivative
\begin{equation}
 \gamma  _m\equiv \frac{m}{m_0^2}\frac{\partial m_0^2}{\partial m}
=\gamma_\pi-\zeta_\pi,
\label{@gm}\end{equation}
the right-hand side following from (\ref{flowm02}).
By analogy with the other effective quantities
we define
the effective exponent

\begin{equation}
\label{nubar}
\bar{\nu}(\kappa)=\frac{1}{2+\bar{\zeta}_{\pi}(\kappa)-
\bar{\eta}(\kappa)},
\end{equation}
where $\bar{\zeta}_{\pi}(\kappa)=\zeta_\pi(\bar{f}(\kappa))$.
The effective exponent $\bar{\nu}(\kappa)$ gives the
critical exponent $\nu$ at the fixed point  $\kappa=\kappa_*$.
To leading order in $\Delta\kappa$ we obtain

\begin{equation}
\label{nubar1}
\bar{\nu}(\kappa)=
\frac{1}{2}+\frac{165}{\sqrt{2}}\Delta\kappa^3.
\end{equation}

By substituting
the previously found fixed-point value
 $\kappa_*=1.17/\sqrt{2}$
of Eq.~(\ref{@fpvs}) into (\ref{nubar}), we
obtain $\nu\approx 1.02$. To estimate the systematic error,
we calculate the fixed-point value $ \kappa ^*$
from the expanded $\beta$-function
(\ref{betak2expansion}), which yields $\kappa_*=1.22/\sqrt{2}$.
Inserting this into Eq. (\ref{nubar1})
gives $\nu\approx 0.92$. The expanded  Eq. (\ref{etaexpansion}) yields
 $\eta\approx -0.03$.

The critical values of $ \nu $ are quite far from the expected XY-model value
$ \nu _{XY}\approx0.67$, but
at the one-loop level
we should not expect a higher accuracy.

Systematic improvements in our approach is considerably more difficult than 
in other more conventional approaches. The fundamental difference 
between our method and other methods lies in the fact that we perturbatively 
expand in powers of $f$ only. The essence of the method lies in the 
computation of $\gamma_A$ and further determination of 
$\bar{f}$ using Eq. (\ref{fbareq}). Generally we have 
that $\gamma_A$ is given by the series:

\begin{equation}
\label{gammaAexp}
\gamma_A(f,\kappa)=\sum_{l=1}^\infty c_l(\kappa) f^l.
\end{equation}
The powers of $f$ correspond to the number of loops and the 
coefficients $c_l(\kappa)$ are not polynomials in $\kappa$ and 
should diverge at some value $\kappa=\kappa_t$ separating 
type I from type II regimes. At one-loop order $\kappa_t=1/\sqrt{2}$, 
receiving no corrections with respect to the mean-field value. 

\subsection{Scaling behavior as $T_c$ is approached from below}

Let us discuss the scaling behavior as $T_c$ is approached from 
below from a perspective independent of perturbation theory. It is 
convenient to write the formulas for any dimension $d\in(2,4)$. Thus, 
the $\beta$-funtion for the gauge coupling is given by

\begin{equation}
\label{betafd}
\beta_f=(\gamma_A+d-4)f.
\end{equation}
The $\beta$-function $\beta_{\kappa^2}$ can be written as 

\begin{equation}
\beta_{\kappa^2}=\left(4-d-\gamma_A+\frac{\beta_g}{g}\right)\kappa^2.
\end{equation}
Since $m_A^2=m^2/\kappa^2$, we obtain

\begin{equation}
\label{mARGeq}
m\frac{\partial m_A^2}{\partial m}=\left(d-2+\gamma_A-
\frac{\beta_g}{g}\right)m_A^2.
\end{equation}
The existence of a charged fixed point implies 
$\eta_A\equiv\gamma_A^*=4-d$ and $\beta_g^*=0$. Thus, in the 
neighborhood of this charged fixed point Eq. (\ref{mARGeq}) 
becomes $m\partial m_A^2/\partial m\approx 2m_A^2$, implying 
that $m_A^2\sim m^2$. Therefore, $\lambda$ and $\xi$ diverge 
with the same critical exponent \cite{Herbut,Olsson}. 
This is of course obvious from 
the definition of the Ginzburg parameter: 
$\kappa^2=g/2f=\lambda^2/\xi^2$. If $g_*$ and $f_*$ are different from 
zero the ratio $\lambda/\xi$ is a constant at the critical point. This 
can only happens if they have the same exponent. At the $XY$ fixed 
point, however, $\eta_A=0$ because $f_*=0$ and 
$m\partial m_A^2/\partial m\approx(d-2)m_A^2$ near the fixed point. 
In this case $m_A^2\sim m^{d-2}$ and the penetration depth 
exponent is given by $\nu'=\nu(d-2)/2$. This is the so called 
$XY$ superconducting universality class \cite{Fisher}. The 
universality class governed by the charged fixed point is 
fundamentally different from the $XY$ universality class. 
Note that the above argument works only if we approach $T_c$ 
from below, since $m_A^2\neq 0$ only for $T<T_c$.

From Eqs. (\ref{mA2}) and (\ref{superfdens}) we can write 
$m_A^2=e^2\rho_s$. Due to Eq. (\ref{betafd}),  $e^2\sim m^{4-d}$ 
near the critical point governed by the charged fixed point. Since 
there $m_A^2\sim m^2$, we obtain the Josephson relation 
\cite{Josephson} $\rho_s\sim m^{d-2}$.

\section{Momentum space instabilities. Origin of the
negative sign of the $\eta$-exponent}

A much debated property of the superconducting phase transition is
the sign of the $\eta$-exponent
\cite{deCalan1,Nogueira1,Nguyen,Hove,Nogueira2}. Recently, it was
pointed out by one of us\cite{Nogueira2} (F.S.N.) that the
origin of the negative sign of $\eta$ lies on momentum
space instabilities of the order-parameter correlation function
arising from
 magnetic field fluctuations.
Such momentum space instabilities are very similar to those
occuring in scalar models of Lifshitz points\cite{Hornreich}.
In these models, the $\eta$-exponent (which in this context is
usually denoted by
$\eta_{l4}$) is also found to be negative. The only
difference
with respect to the GL model is that
in the  scalar models for Lifshitz points
the momentum space instabilities are
included from the beginning explicitly into the Hamiltonian.

Inspired by works on lattice spin models of
Lifshitz points\cite{Selke},
we expand $\Gamma_{0,\pi\pi}^{(2)}$ up to order $({\bf p}^2)^3$:

\begin{equation}
\label{expansion1}
\Gamma_{0,\pi\pi}^{(2)}({\bf p})\simeq{\bf p}^2\left[Z_\pi^{-1}
+\frac{Af}{\pi}\frac{{\bf p}^2}{m_A^2}+
\frac{Bf}{\pi}\left(\frac{{\bf p}^2}{m_A^2}\right)^2\right],
\end{equation}
where

\begin{equation}
\label{A} \!\!\!
A=\frac{\sqrt{2}}{4}
(1-\sqrt{2}\kappa)
(\sqrt{2}\kappa+1)^{-3}
\left(\frac{1}{6}+\frac{2}{5}\sqrt{2}\kappa+\frac{
\kappa^2}{5}\right),
\end{equation}

\begin{equation}
B=\frac{1}{56\sqrt{2}}-\frac{1}{105\sqrt{2}}
\frac{1\!+\!5\sqrt{2}\kappa
\!+\!20\kappa^2\!+\!18\sqrt{2}\kappa^3}{(\sqrt{2}\kappa+1)^{5}}.
 \label{@B}
\end{equation}
The above expansion is like a Landau expansion where ${\bf p}$
play the role of an ``order parameter''.
Remarkably, the coefficient $A$ vanishes for $\kappa=1/\sqrt{2}$.
In the type I regime $A$ has a positive sign while in the
type II regime $A$ is negative. The coefficient $B$ is
always positive, ensuring the stability.
Close to $\kappa=1/\sqrt{2}$ we find

\begin{equation}
A\simeq-\frac{1}{24}\left(\kappa-\frac{1}{\sqrt{2}}\right)
=-\frac{\Delta\kappa}{24},
\end{equation}

\begin{equation}
B\simeq\frac{1}{96\sqrt{2}}.
\end{equation}
For $\kappa\gtsim 1/\sqrt{2}$, $\Gamma_{0,\pi\pi}^{(2)}({\bf p})$
has a zero at

\begin{equation}
\label{p0}
|{\bf p}_0|\simeq 2^{3/4}m_A\Delta\kappa^{\bar{\beta}},
\end{equation}
with the exponent $\bar{\beta}=1/2$. Note that
$\Gamma_{0,\pi\pi}^{(2)}({\bf p})$ is just the inverse of the transverse
susceptibility $\chi_T({\bf p})$. Thus, the transverse
susceptibility is maximized at $|{\bf p}_0|$. This is the behavior
assumed  so far
in all generic models exhibiting a
Lifshitz point\cite{Hornreich}. It
implies the existence of a
modulated regime for the order parameter. On the basis
of these considerations we conclude
 that the transition from
type I to type II behavior happens at a Lifshitz point.

This conclusion is, in fact, in accordance
with another physical property of the
transition from type I to type II superconductors.
In the second, fluctuating vortex lines play an important role.
This is precisely the reason why
the fluctuation term
$\sim|\psi|^{4-\epsilon}$
calculated under the assumption of a constant $|\psi|$
is unreliable.
At the core of each vortex line,
$|\psi|$ has to vanish, such that
in the type II regime the order field os perforated by lines of zeros
\cite{rem1}.

In our one-loop calculation the  $\kappa$-value
where a type I superconductor
crosses over to type II
has the mean-field values $\kappa_L=1/\sqrt{2}$.
This value is expected to decrease  in higher-loop calculations
towards the tricritical value $ \kappa _t\approx0.77/ \sqrt{2}$
where we expect the cross over to take place.
Indeed, in the dual theory
\cite{Kleinert,KleinertBook}, the tricritical point
comes about by the sign change of the quartic term
in the disorder field
which corresponds
to a change of the average
interaction between vortex lines changing from repulsion to attraction.
This sign is also what distinguishes the type I from the type II regime
experimentally.

 It remains to be checked by
going to higher loops, that
the fluctuation corrected value of $\kappa_L$
does indeed coincide
with the
Ginzburg parameter at the tricritical point, that is,
$\kappa_L\simeq 0.77/\sqrt{2}$ \cite{Kleinert,SUD}.

The exact value of the exponent
$\bar{\beta}$ controlling the vanishing of $|{\bf p}_0|$ at the type-I-type II
boundary
in Eq.~(\ref{p0})
is expected to increase slightly above $1/2$.

An  interesting  open problem is
to
prove the conjecture
of Ref. \onlinecite{Nogueira2}
that the Lifshitz point
coincides with the tricritical point obtained from the disorder
field theory \cite{Kleinert}.

\section{Conclusion}

In this paper we have initiated a new RG approach to the GL model.
Since we work directly in $d=3$ dimensions
 and in the ordered phase,
we were able to express the RG functions as functions
of the Ginzburg parameter $\kappa=m/m_A$. This replaces the
coupling constant
$g$ used above $T_c$, since we can also write $\kappa^2=g/2f$.
The  RG functions have no longer
a polynomial form in $\kappa$, being polynomial only in
the gauge coupling constant $f$.
The effective gauge coupling $\bar{f}(\kappa)$ determined by
the condition $\beta_f=0$ shows that the charge is small for
$\kappa$ sufficiently close to $\kappa=1/\sqrt{2}$ from
above. Interestingly, the fixed point $\kappa_*$ lies
close to $\kappa=1/\sqrt{2}$.

We have  compared our values for the critical exponents with those
found in the literature.
 Our  one-loop  value of $\nu$ is larger than the $XY$
value  obtained
from Monte Carlo simulations \cite{Olsson} and
from RG analysis of the disorder field theory \cite{Kiometzis2}. Using the
expansion in powers of
$\Delta\kappa$ of Section IV we have obtained
$\eta\approx -0.03$ and $\nu\approx 0.92$. With these values of
$\eta$ and $\nu$ we obtain
a
critical exponent $\gamma=\nu(2-\eta)\approx 1.87$.
This
agrees with the Pad\'e approximant analysis
of the two-loop $\epsilon$-expansion obtained
by Folk and Holovatch \cite{Folk}. On the basis of
the Pad\'e approximants, these authors obtained the
value $\nu\approx 0.857$, showing a $\sim 7\%$ numerical
difference from our result $\nu\approx 0.92$.
The exponent $\gamma$ obtained from recent Monte Carlo simulations is
$\gamma\approx 1.45$ \cite{Nguyen}.

One question that immediately arises is how the critical behavior obtained 
with our method is related to approaches where $T_c$ is approached 
from above. In the pure $O(N)$ 
symmetric $\phi^4$ theory the critical singularities  
above and below $T_c$ are known to be the same. This result is 
obtained by analysing the Ward identities of the theory. For a 
dimensionally regularized theory, the same counterterms used in the 
disordered phase also renormalize the theory in the ordered phase 
\cite{ZJ}. Since the critical exponents in the $\epsilon$-expansion 
are a direct consequence of the $1/\epsilon$  singularities 
included in the counterterms, we conclude that the critical 
exponents obtained by approaching $T_c$ from below are the same 
as when $T_c$ is approached from above. This should be true even 
if we work in fixed dimension. If we use the $\epsilon$-expansion 
in the GL model, the same argument holds. The problem in this case is that 
the superconducting fixed point appears only for sufficiently large 
$N$. We have seen that the value of $N$ can be dramatically reduced 
if we consider a fixed dimension approach for $T\geq T_c$. 
Unfortunately this is not enough to obtain a charged fixed point 
for $N=2$. Then we have shown that by approaching $T_c$ from below 
we reach a charged fixed point for the physically interesting 
$N=2$ case. It is not obvious which renormalization scheme 
above $T_c$ corresponds to our scheme below $T_c$. Clearly, 
by working above $T_c$ we would never find the same 
renormalization constants. 
The point is that $\kappa$, which appears 
explicitly below $T_c$ in a non-polynomial form, can be 
only defined above $T_c$ through the ratio between the coupling 
constants. The vector potential is massive only below $T_c$.

We have discussed in detail the agreement
of our theoretical
$\eta$-exponent with recent Monte Carlo
simulations.
It is now definite
that it is negative. The only possible way to enforce
a positive $\eta>0$ in the GL model is by adding an
appropriate mass term,
 like the Proca term considered
in Ref. \onlinecite{Nogueira1}. This should not
 be confused with
our calculation of Section II where $M\to 0$ at the end.
A negative $ \eta $ is caused by the attraction of the vortex lines
on the average which causes a fluctuating vortex
globule to have a smaller size
than  a free random chain. This point of view is related
to the geometric interpretation considered in
Ref. \onlinecite{Mo}.
The addition of a
Chern-Simons term which changes the size of vortex globules
by a topological interaction seems incapable of producing a
positive $ \eta $,
 since entanglement of vortex lines
also tends to contract the globules.
Indeed,
one-loop calculation gives a negative $\eta$ for all values
of the topological coupling \cite{KleinertCS,deCalanCS}.

There is, however, the possibility that
for larger topological coupling $ \eta $ could become positive,
since for infinite topological coupling $\eta\to 0$.
This is suggested when calculating  $\eta$ in
 a $1/N$ expansion \cite{KleinNog}.

We hope that the new approach introduced in this paper will
stimulate
further discussions of the nature of the superconducting
phase transition. In particular,
we call for two- and higher-loop calculations
to substantiate
our claim
that $ \kappa _L= \kappa _t$, and that the type-I-type-II
crossover point is a Lifshitz point.

\acknowledgments

The authors would like to thank B. Van den Bossche and
Prof. A. Sudb{\o} for many interesting discussions. They
thank in particular A. M. J. Schakel for his very
enlightening comments. The research of FSN is
supported by the Alexander von Humboldt foundation.

\appendix\section{Useful integrals}

In this appendix we shall write some basic
momentum space integrals in three dimensions that are used
to obtain the results in the text.

The simplest integral, arising when computing tadpoles, is

\begin{equation}
\int\frac{d^3k}{(2\pi)^3}\frac{1}{{\bf k}^2+m^2}=
\frac{\Lambda}{2\pi^2}-\frac{m}{4\pi}+{\cal O}(m/\Lambda),
\end{equation}
where the ultraviolet cutoff is assumed to be very large,
$\Lambda\gg m$.

The following type of integral appears in the
calculation of the two-point functions:

\begin{equation}
\int\frac{d^3k}{(2\pi)^3}\frac{1}{[({\bf p}-{\bf k})^2+m_1^2]
({\bf k}^2+m_2^2)}=\frac{1}{8\pi|{\bf p}|}a({\bf p},m_1,m_2),
\end{equation}
where $a({\bf p},m_1,m_2)$ is defined in Eq. (\ref{a}).
The following particular cases of the above integral are relevant:

\begin{equation}
\int\frac{d^3k}{(2\pi)^3}\frac{1}{[({\bf p}-{\bf k})^2+m^2]
({\bf k}^2+m^2)}=\frac{1}{8\pi|{\bf p}|}b({\bf p},m),
\end{equation}
where $b({\bf p},m)$ is defined in Eq. (\ref{b}), and

\begin{equation}
\int\frac{d^3k}{(2\pi)^3}\frac{1}{({\bf k}^2+m_1^2)
({\bf k}^2+m_2^2)}=\frac{1}{4\pi(m_1+m_2)}.
\end{equation}

When computing loop integrals involving the vector potential
propagator we need the integrals:

\begin{equation}
\int\frac{d^3k}{(2\pi)^3}\frac{k_\mu}{[({\bf p}-{\bf k})^2+m_1^2]
({\bf k}^2+m_2^2)}=\frac{p_\mu}{16\pi|{\bf p}|^3}[2|{\bf p}|(
m_2-m_1)+({\bf p}^2+m_1^2-m_2^2)a({\bf p},m_1,m_2)],
\end{equation}

\begin{eqnarray}
&&\int\frac{d^3k}{(2\pi)^3}\frac{k_\mu k_\nu}{[({\bf p}-{\bf k})^2+m_1^2]
({\bf k}^2+m_2^2)}=-\frac{\delta_{\mu\nu}}{32\pi}
\left\{m_1+m_2-\frac{(m_1+m_2)(m_1-m_2)^2}{{\bf p}^2}
\right.\nonumber
\\*[0.8cm]
&+&\left.
\frac{[{\bf p}^2+(m_1+m_2)^2][{\bf p}^2+(m_1-m_2)^2]}{2|{\bf p}|^3}
a({\bf p},m_1,m_2)\right\}
+\frac{p_\mu p_\nu}{32\pi{\bf p}^2}\left\{3m_2-5m_1
\right.\nonumber
\\*[0.8cm]
&-&\left.\frac{3(m_1+m_2)(m_1-m_2)^2}{{\bf p}^2}
+\frac{3({\bf p}^2+m_1^2)^2-6m_1^2m_2^2+3m_2^2-2m_2^2{\bf p}^2}{
2|{\bf p}|^3}a({\bf p},m_1,m_2)\right\}.
\end{eqnarray}

\newpage
\begin{figure}
\setlength{\unitlength}{1mm}
\begin{fmffile}{graph4p}

\beq
\parbox{2cm}{
\centerline{
  \begin{fmfgraph}(25,20)
  \setval
  \fmfforce{0w,0.8h}{v1}
  \fmfforce{0w,0.2h}{v2}
  \fmfforce{1/4w,1/2h}{v3}
  \fmfforce{3/4w,1/2h}{v4}
  \fmfforce{1w,0.8h}{v5}
  \fmfforce{1w,0.2h}{v6}
  \fmf{plain,width=0.3mm}{v1,v3,v2}
  \fmf{plain,width=0.3mm}{v5,v4,v6}
  \fmf{wiggly,width=0.3mm,left=1}{v3,v4,v3}
  \fmfdot{v3,v4}
  \end{fmfgraph}}}\nonumber
\eeq
\end{fmffile}
\caption{Feynman graph contributing to the four-point function.
The external lines correspond to the scalar fields, while
the wiggles are the vector potential propagators.}
\end{figure}
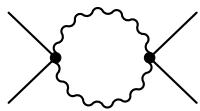

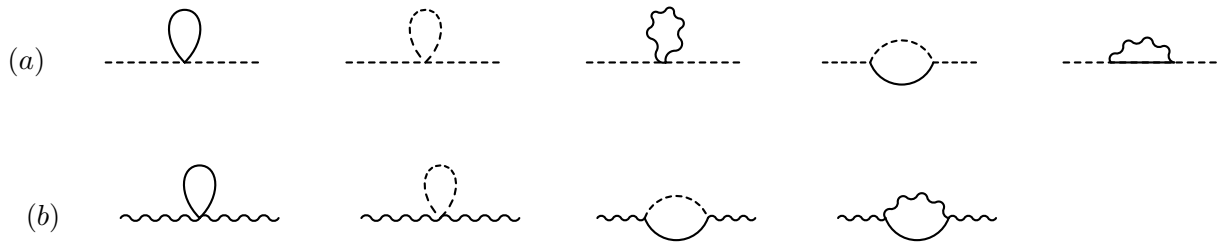
\begin{figure}

\setlength{\unitlength}{0.7mm}
\begin{fmffile}{graph1}
\beq
 ~(a)~~
\parbox{32mm}{
\centerline{
\begin{fmfgraph}(30,10)
\setval
\fmfforce{0w,1/2h}{v1}
\fmfforce{1/2w,1/2h}{v2}
\fmfforce{1w,1/2h}{v3}
\fmf{dashes,width=0.3mm}{v1,v2,v3}
\fmf{plain,width=0.3mm}{v2,v2}
\end{fmfgraph}}}
\hspace*{0.cm}
\parbox{32mm}{
\centerline{
\begin{fmfgraph}(30,10)
\setval
\fmfforce{0w,1/2h}{v1}
\fmfforce{1/2w,1/2h}{v2}
\fmfforce{1w,1/2h}{v3}
\fmf{dashes,width=0.3mm}{v1,v2,v3}
\fmf{dashes,width=0.3mm}{v2,v2}
\end{fmfgraph}}}
\hspace*{0.cm}
\parbox{32mm}{
\centerline{
\begin{fmfgraph}(30,10)
\setval
\fmfforce{0w,1/2h}{v1}
\fmfforce{1/2w,1/2h}{v2}
\fmfforce{1w,1/2h}{v3}
\fmf{dashes,width=0.3mm}{v1,v2,v3}
\fmf{wiggly,width=0.3mm}{v2,v2}
\end{fmfgraph}}}
\hspace*{0.cm}
\parbox{32mm}{
\centerline{
  \begin{fmfgraph}(30,10)
  \setval
  \fmfforce{0w,1/2h}{v1}
  \fmfforce{1w,1/2h}{v2}
  \fmfforce{9/30w,1/2h}{v3}
  \fmfforce{21/30w,1/2h}{v4}
  \fmf{dashes,width=0.3mm}{v1,v3}
  \fmf{dashes,width=0.3mm}{v4,v2}
  \fmf{plain,width=0.3mm,right=0.7}{v3,v4}
  \fmf{dashes,width=0.3mm,left=0.7}{v3,v4}
  \end{fmfgraph} } }
\hspace*{0.cm}
  \parbox{32mm}{\centerline{
  \begin{fmfgraph}(30,10)
  \setval
  \fmfforce{0w,1/2h}{v1}
  \fmfforce{1w,1/2h}{v2}
  \fmfforce{9/30w,1/2h}{v3}
  \fmfforce{21/30w,1/2h}{v4}
  \fmf{dashes,width=0.3mm}{v1,v2}
  \fmf{plain,width=0.3mm}{v3,v4}
  \fmf{wiggly,width=0.3mm,left=0.7}{v3,v4}
  \end{fmfgraph} } }  \nonumber
\\*[1.2cm]
\!\!\!\!\!\!(b)~~
\parbox{32mm}{
\centerline{
\begin{fmfgraph}(30,10)
\setval
\fmfforce{0w,1/2h}{v1}
\fmfforce{1/2w,1/2h}{v2}
\fmfforce{1w,1/2h}{v3}
\fmf{wiggly,width=0.3mm}{v1,v2,v3}
\fmf{plain,width=0.3mm}{v2,v2}
\end{fmfgraph}}}
\hspace*{0.cm}
\parbox{32mm}{
\centerline{
\begin{fmfgraph}(30,10)
\setval
\fmfforce{0w,1/2h}{v1}
\fmfforce{1/2w,1/2h}{v2}
\fmfforce{1w,1/2h}{v3}
\fmf{wiggly,width=0.3mm}{v1,v2,v3}
\fmf{dashes,width=0.3mm}{v2,v2}
\end{fmfgraph}}}
\hspace*{0.cm}
\parbox{32mm}{\centerline{
  \begin{fmfgraph}(30,10)
  \setval
  \fmfforce{0w,1/2h}{v1}
  \fmfforce{1w,1/2h}{v2}
  \fmfforce{9/30w,1/2h}{v3}
  \fmfforce{21/30w,1/2h}{v4}
  \fmf{wiggly,width=0.3mm}{v1,v3}
  \fmf{wiggly,width=0.3mm}{v4,v2}
  \fmf{plain,width=0.3mm,right=0.7}{v3,v4}
  \fmf{dashes,width=0.3mm,left=0.7}{v3,v4}
  \end{fmfgraph} } }
\hspace*{0.cm}
  \parbox{32mm}{\centerline{
  \begin{fmfgraph}(30,10)
  \setval
  \fmfforce{0w,1/2h}{v1}
  \fmfforce{1w,1/2h}{v2}
  \fmfforce{9/30w,1/2h}{v3}
  \fmfforce{21/30w,1/2h}{v4}
  \fmf{wiggly,width=0.3mm}{v1,v3}
  \fmf{wiggly,width=0.3mm}{v4,v2}
  \fmf{plain,width=0.3mm,right=0.7}{v3,v4}
  \fmf{wiggly,width=0.3mm,left=0.7}{v3,v4}
  \end{fmfgraph} } }\phantom{cccccccccccccccccc}
\nonumber \eeq
\end{fmffile}
\caption{Graphs contributing to the evaluation of the
$\pi$ (a) and vector potential (b) two-point functions.
The solid lines indicate the $\sigma$-propagators,
the dashed ones $\pi$-propagators.
The wiggles represent the vector potential.
}

\end{figure}

\vspace{1cm}

\begin{figure}[tbhp]
\begin{picture}(0,84)
\unitlength.7mm
\put(10,-25){\includegraphics[width=3cm]{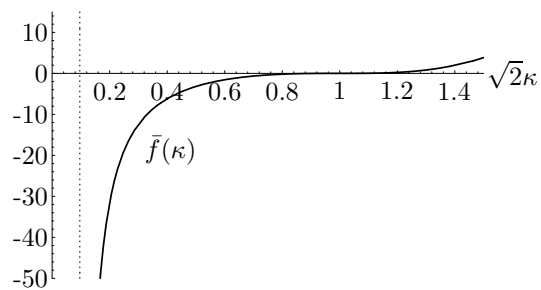}}
\put(35,18){\fsz$ \bar f( \kappa ) $}
\put(100,32){\fsz$ \sqrt{2} \kappa $}
 \end{picture}
\caption[]{Behavior of effective square charge $\bar f( \kappa )$.
To the left of the vertical dashed line
marking the
the interval $0\leq \sqrt{2} \kappa<0.096$,
$\bar{f}( \kappa )$ is very large and positive, starting out from
$\bar{f}(0)\approx53.3$ and going to infinity for
$ \kappa \rightarrow0.096/\sqrt{2} $.
   }
\label{fbarfig}
\end{figure}

\begin{figure}
{}~~~~~~\begin{picture}(87.87,143.05)
\unitlength.7mm
\put(-18,3){\includegraphics{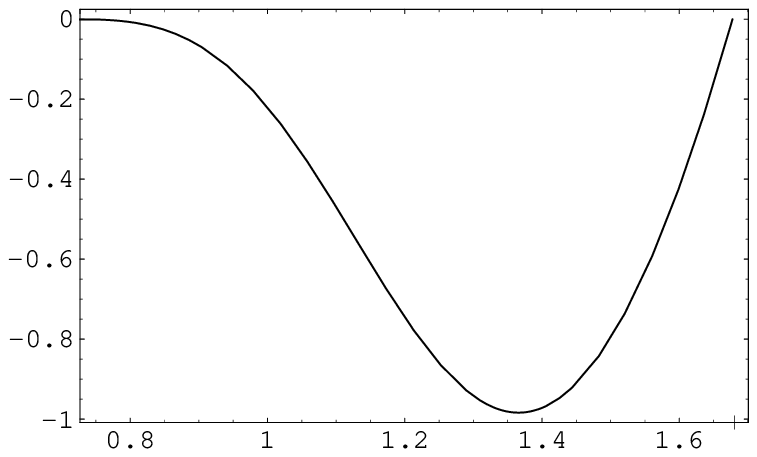}
}
\put(-3,6.){\fsz$1/ \sqrt{2}  $}
\put(58,6.5){\fsz$ \kappa $}
\put(95.9,6.5){\fsz$\hat \kappa $}
\put(27,47.5){\fsz$\bar \eta( \kappa )$}
\end{picture}
\caption[]{Plot of $\bar{\eta}$ as a function of $\kappa$ in the
interval $1/\sqrt{2}<\kappa<\hat  \kappa \equiv (\sqrt{33}-1)/2\sqrt{2}
\approx1.67746$.
{}}
\label{f2}
\end{figure}


\begin{thebibliography}{100}

\bibitem{ZJ} J. Zinn-Justin, {\it Quantum Field Theory and
Critical Phenomena} (Oxford, 1993).

\bibitem{HKSF}
  H. Kleinert and V. Schulte-Frohlinde, {\em
     Critical Phenomena in $\Phi ^4$-Theory\/},
     World Scientific, Singapore 2001
(http://www.physik.fu-berlin.de/{}\~{}{}kleinert/re0.html\#b8)


\bibitem{HLM1} B. I. Halperin, T. C. Lubensky and S.-K. Ma,
Phys. Rev. Lett. {\bf 32}, 292 (1974); J.-H. Chen, T. C. Lubensky
and D. R. Nelson, Phys. Rev. B {\bf 17}, 4274 (1978)

\bibitem{CW}
S. Coleman and E. Weinberg, Phys. Rev. D {\bf 7}, 1888 (1983).
For a comparison of the two theories
see
H. Kleinert,
Phys. Lett. B {\bf 128}, 69 (1983).


\bibitem{Schakel} For a recent discussion
of this subject see
A. M. J. Schakel, ``Effective free energy of
Ginzburg-Landau model'' in: {\it Fluctuating Paths and Fields},
Eds. W. Janke, A. Pelster, H.-J. Schmidt, and M. Bachmann
(World Scientific, Singapore, 2001).


\bibitem{Dasgupta} C. Dasgupta and B. I. Halperin, Phys.
Rev. Lett. {\bf 47}, 1556 (1981).

\bibitem{LawAth} I. D. Lawrie and C. Athorne, J. Phys. A {\bf 16},
L587 (1983).

\bibitem{AN} J. Als-Nielsen et al., Phys. Rev. B {\bf 22}, 312 (1980).

\bibitem{deGennes} P.-G. de Gennes, Solid State Commun. {\bf 10}, 753
(1972); B. I. Halperin and T. C. Lubensky, {\it ibid.} {\bf 14}, 997
(1974).


\bibitem{Kleinert} H. Kleinert, Lett. Nuovo Cimento {\bf 35}, 405 (1982).

\bibitem{KleinertBook} For details of the
derivation see Chapter 13 in
the
textbook by H. Kleinert, {\it Gauge Fields in Condensed
Matter, vol. 1}, (World Scientific, Singapore, 1989),
readable in the internet at
http://www.physik.fu-berlin.de/\~{}kleinert/re.html\#b1.

\bibitem{Kiometzis2} M. Kiometzis, H. Kleinert and A. M. J.
Schakel, Phys. Rev. Lett. {\bf 73}, 1975 (1994)
(cond-mat/9503019);
Fortschr. Phys. {\bf 43}, 697 (1995)
(cond-mat/9508142).

\bibitem{Cam}
 H.~Kleinert,
    {\em Theory of Fluctuating Nonholonomic Fields and Applications:
     Statistical Mechanics of Vortices and Defects and New Physical
      Laws in Spaces with Curvature and Torsion\/},
     in: Proceedings of NATO Advanced Study Institute on
       Formation and Interaction of Topological Defects
at the University of Cambridge,
     Plenum Press, New York, 1995, S.~201--232 (cond-mat/9503030)
%



\bibitem{SUD}
S. Mo, J. Hove, A. Sudb\o, Phys. Rev. B {\bf 65}, 104501 (2002).

\bibitem{Olsson} P. Olsson and S. Teitel, Phys. Rev. Lett. {\bf 80},
1964 (1998).


\bibitem{Tess} J. Tessmann, MS thesis 1984
written  under the supervision of one
the present authors (HK); the pdf file is available in the
internet http://www.physik.fu-berlin.de/\~{}kleinert/MS-Tessmann.pdf.

\bibitem{Kolnberger} S. Kolnberger and R. Folk, Phys. Rev. B
{\bf 41}, 4083 (1990).

\bibitem{NoteTess} There is a slight mistake in 
Tessmann's calculation at the two-loop level for the 
$\beta$ function of the $g$ coupling.   

\bibitem{Folk} R. Folk and Y. Holovatch, J. Phys. A {\bf 29}, 3409 (1996).

\bibitem{Kast} B. Kastening, H. Kleinert, and B. Van den Bossche,
Phys. Rev. B {\bf 65}, 174512 (2002).

\bibitem{Berg} B. Bergerhoff, F. Freire, D. F. Litim, S. Lola and
C. Wetterich, Phys. Rev. B {\bf 53}, 5734 (1996).

\bibitem{ERG} K. G. Wilson and Kogut, Phys. Rep. {\bf 12}, 
75 (1974); F. J. Wegner, in ``Phase Transitions and 
Critical Phenomena'', Vol. 6, eds. C. Domb and M. S. Green 
(Academic Press, New York, 1976); J. Polchinski, 
Nucl. Phys. B {\bf 231}, 269 (1984); C. Wetterich, 
Phys. Lett. B{\bf 301}, 90 (1993).

\bibitem{Herbut} I. F. Herbut and Z. Te\u{s}anovi\'c,
Phys. Rev. Lett. {\bf 76}, 4588 (1996).

\bibitem{Barth} J. Bartholomew, Phys. Rev. B {\bf 28}, 5378 (1983).

\bibitem{Mou} C.-Y. Mou, Phys. Rev. B {\bf 55}, R3378 (1997).

\bibitem{Nogueira2} F. S. Nogueira, Phys. Rev. B {\bf 62}, 14559 (2000).

\bibitem{Fisher} D. S. Fisher, M. P. A. Fisher and D. A. Huse,
Phys. Rev. B {\bf 43}, 130 (1991).

\bibitem{Chen} G.-H. Chen and Y.-S. Wu, hep-th/0110134.

\bibitem{Parisi} G. Parisi, J. Stat. Phys. {\bf 23}, 49 (1980).

\bibitem{Lawrie} I. D. Lawrie, Nucl. Phys. B {\bf 200}, 1 (1982).

\bibitem{Nogueira1} F. S. Nogueira, Europhys. Lett. {\bf 45}, 612 (1999).


\bibitem{deCalan1} C. de Calan and F. S. Nogueira, Phys. Rev. B {\bf 60},
4255 (1999).

\bibitem{Nguyen} A. K. Nguyen and A. Sudb\o, Phys. Rev. B {\bf 60},
15307 (1999).

\bibitem{Hove} J. Hove and A. Sudb{\o}, Phys. Rev. Lett {\bf 84}, 3426 (2000).

\bibitem{Olsson1} P. Olsson, e-print: cond-mat/0103345.

\bibitem{Note1} In the $1/N$ expansion it is convenient to rescale
the couplings $e_0\to e_0/\sqrt{N}$ and $u_0\to u_0/N$.

\bibitem{Lee} B. W. Lee, Phys. Rev. D {\bf 5}, 823 (1972).

\bibitem{Josephson} B. D. Josephson, Phys. Lett. {\bf 21}, 608 (1966).

\bibitem{Grunberg} G. Grunberg, e-print: hep-ph/0009272, and
references therein.

\bibitem{Hornreich} R. M. Hornreich, M. Luban and S. Shtrikman,
Phys. Rev. Lett. {\bf 35}, 1678 (1975);
R. M. Hornreich, M. Luban and S. Shtrikman,
Phys. Lett. A {\bf 55}, 269 (1975).

\bibitem{Selke} W. Selke, in {\it Phase Transitions and Critical
Phenomena}, edited by C. Domb and J. L. Lebowitz
(Academic London, 1992), Vol. 15, pp. 1-72.

\bibitem{Abrikosov} A. A. Abrikosov, Soviet Phys. (JETP) {\bf 5},
1174 (1957).

\bibitem{Radz} L. Radzihovsky, Europhys. Lett. {\bf 29}, 227 (1995).

\bibitem{KleinertCS} H. Kleinert and A. M. J. Schakel, FU-Berlin
preprint, 1993 (unpublished)
(http://www.physik.fu-berlin.de/\~{}kleinert/215).

\bibitem{deCalanCS} C. de Calan, A. P. C. Malbouisson, F. S. Nogueira and
N. F. Svaiter, Phys. Rev. B {\bf 59}, 554 (1999).

\bibitem{deCalan3} C. de Calan and F. S. Nogueira, Phys. Rev. B
{\bf 60}, 11929 (1999).

\bibitem{King} T. Kennedy and C. King, Phys. Rev. Lett. {\bf 55}, 776
(1985); Commun. Math. Phys. {\bf 104}, 327 (1986); C. Borgs and
F. Nill, {\it ibid.} {\bf 104}, 349 (1986).

\bibitem{Spencer} J. Fr\"ohlich, B. Simon and T. Spencer, Commun. Math.
Phys. {\bf 50}, 79 (1976).

\bibitem{Kiometzis1} M. Kiometzis and A. M. J. Schakel,
Int. J. Mod. Phys. B {\bf 7}, 4271 (1993).

\bibitem{tHooft} G. 't Hooft, Nucl. Phys. B {\bf 35}, 167 (1971).

\bibitem{Mo} J. Hove, S. Mo, and A. Sudb{\o}, Phys. Rev. Lett. {\bf 85},
2368 (2000).


\bibitem{KleinNog} H. Kleinert and F. S. Nogueira, J. Phys. Stud. 
{\bf 5}, 327 (2001).

\bibitem{rem1}
This
phenomenon
is discussed in detail in the textbook Ref.
\onlinecite{KleinertBook}.

\bibitem{Collins} J. C. Collins, {\it Renormalization}
(Cambridge University Press, Cambridge, 1984).

\end{thebibliography}
\end{document}